\documentclass[aps,twocolumn,superscriptaddress,floatfix,showpacs]{revtex4}
\usepackage{graphicx,amssymb,amsmath}
\newcommand{\form}{\mathrm{C}_N \mathrm{H}_2}
\newcommand{\ctwentyone}{\mathrm{C}_{21} \mathrm{H}_2}
\newcommand{\cfourtyone}{\mathrm{C}_{41} \mathrm{H}_2}
\newcommand{\csixtyone}{\mathrm{C}_{61} \mathrm{H}_2}
\newcommand{\chundred}{\mathrm{C}_{100} \mathrm{H}_2}
\newcommand{\chundredone}{\mathrm{C}_{101} \mathrm{H}_2}
\newcommand{\chundredfourtyone}{\mathrm{C}_{141} \mathrm{H}_2}
\newcommand{\chundredfourty}{\mathrm{C}_{140} \mathrm{H}_2}
\newcommand{\dn}{\delta_n}
\newcommand{\dno}{\delta_0}
\newcommand{\qn}{q_n}
\newcommand{\rn}{\rho_n^s}
\newcommand{\Dx}{\Delta (x)}

\newcommand{\Do}{\Delta_0}
\newcommand{\kso}{\xi_0}
\newcommand{\Etot}{E_{\mathrm{tot}}}
\newcommand{\EA}{\mathrm{EA}}

\begin{document}

\title{A comparative \textit{ab initio} study of neutral and charged kink-solitons on conjugated carbon chains}

\author{M.~L.~Mayo}
\affiliation{Department of Physics, The University of Texas at
Dallas, P. O. Box 830688, EC36, Richardson, Texas 75083, USA}
\author{Yu.~N.~Gartstein}
\affiliation{Department of Physics, The University of Texas at
Dallas, P. O. Box 830688, EC36, Richardson, Texas 75083, USA}

\begin{abstract}
The ground state of odd-$N$ polyynic oligomers $\form$ features kink-solitons in carbon-carbon bond-length alternation (BLA) patterns. We perform a systematic first-principles computational study of neutral and singly-charged kinks in long oligomers addressing relationships between BLA patterns, electron energy gaps, and accompanying distributions of spin and charge densities, both in vacuum and in the screening solvent environment. A quantitative comparison is made of the results derived with four different \textit{ab initio} methods: from pure DFT to pure Hartree-Fock (HF) and including two popular hybrid density functionals, B3LYP and BHandHLYP.  A clear correlation is demonstrated between the derived spatial extent of kinks and the amount of HF exchange used in the functionals. For charged kinks, we find a substantial difference in the behavior of charge and spin densities.
\end{abstract}


\maketitle
\section{Introduction}\label{intro}

\begin{figure}
\centering
\includegraphics[scale=0.65]{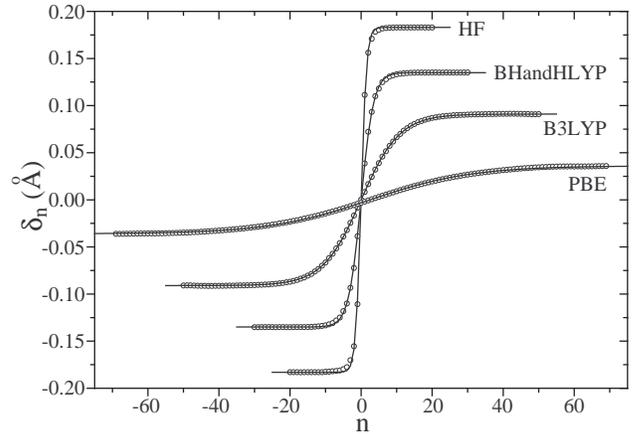}
\caption{Carbon-carbon bond length alternation patterns for neutral triplet kink-solitons on polyynic oligomers in vacuum. The circular data points show results derived from computations with end effects eliminated, solid lines depict analytical fits as discussed in the text. The computational methods used are indicated next to the respective curves.}
\label{fig:BLA_fit}
\end{figure}

One-dimensional (1D) semiconductor nanostructures such as $\pi-$conjugated polymers (CPs), nanotubes and nanowires attract a great deal of attention being both interesting scientifically and important for technological advances.
The nature and properties of excess charge
carriers and electronic excitations on these structures are fundamental for many applications in (opto)electronics, energy harvesting and sensors. As covalent bonds in these systems provide for wide
electronic bands, excess charge carriers
have a high propensity for delocalization and a potential for high
mobilities. The interaction of electrons with other subsystems such as displacements of the underlying atomic lattice and polarization of the surrounding medium can however put inherent limitations on the mobility and may even lead to self-localization of excess carriers.
The electron-lattice coupling has been a subject of a great body of studies for various 1D systems, and in particular for CPs (see, e.g., Refs.~\cite{YuLubook,Bredas1985,Barford_book} for reviews and original references). It is well-known that, as a result of the interaction with lattice distortions,  electrons and holes in CPs may undergo self-trapping into polarons accompanied by localized bond-length modulation patterns and new features in the optical absorption due to local intragap  electronic levels. Kink-solitons are another and qualitatively different type of localized intrinsic ``defects'', which are specific to CPs with the degenerate ground state and exhibit topological kink-like patterns in bond-length modulations (Fig.~\ref{fig:BLA_fit}) and even deeper lying intragap levels. Solitons are believed to play a prominent role in the physics of trans-polyacetylene (for a recent discussion, see Ref.~\cite{lin_1}). Polaronic states, being non-topological excitations, are generic for a broad range of CPs. Polarons in CPs can also be thought of as bound kink-anti-kink pairs.

Numerous first-principles calculations at various theoretical levels have been applied to study the structure of the ground state and excitations in CPs. With such calculations one expects to be able to elucidate effects due to valence electrons and many-electron interactions. Interestingly enough, despite a relatively long history of electron-lattice polarons in CPs (see, e.g., multiple references in  Refs.~\cite{BredasPolOptics,bobbert}), certain questions have been raised recently regarding applicability of different \textit{ab initio} frameworks to describe the polaron formation. That pertains to the reported failure of the local-density-approximation and generalized-gradient-approximation DFT (density-functional theory) schemes to detect self-localized charge density distributions in various charged oligomers, while Hartee-Fock (HF), parameterized semi-empirical and possibly hybrid-functional DFT \cite{Martin} methods have been reported to lead to charge localization in the middle of oligomers (e.g., Refs.~\cite{bobbert,pitea_1,bredas_3,zuppi2003}). Quite different level-of-theory-dependent magnitudes of the effective electron-phonon coupling have also been found for the dimerized ground states of polymeric systems \cite{yang_kertesz_1,yang_kertesz_2}. These observations thus bear on an important general issue of a choice of appropriate \textit{ab initio} methods \cite{Martin,StandardModel}  to faithfully describe properties of 1D semiconductors.

In this paper we contribute to this discussion by comparing results on the structure of kink-solitons in long odd-$N$ polyynic  oligomers $\form$ obtained with different \textit{ab initio} methods including pure HF and DFT (PBE) schemes as well as two popular hybrid-functional  frameworks (B3LYP and BHandHLYP).  Structurally simple polyynic linear carbon chains naturally lend themselves to be a prototypical example of 1D semiconductors.  It should be noted however that polyynic chains continue to be the subject of much attention in their own right \cite{gladysz_1,gladysz_2,yang_kertesz_1,yang_kertesz_2,schaefer_1,schaefer_2}. Interestingly, they were predicted \cite{rice_2,rice_3} to possess a rich family of electron-lattice self-localized excitations because of the extra spatial degeneracy of molecular orbital levels. Neutral kinks take a special, in a sense, place representing in fact the ground-state structure of odd-$N$ oligomers. By studying ``long enough'' oligomers, we are trying to find the ``natural'' spatial extent of the kinks as determined by inherent system interactions rather than by the end effects due to the hydrogen termination. As Fig.~\ref{fig:BLA_fit} illustrates, the spatial extent of solitons turns out to be very strongly dependent on the computational method used and evidently correlated with the magnitude of the equilibrium lattice dimerization.

An addition of an electron (or a hole) to oligomers results in charged kinks and increases their spatial extent. The strong dependence on the computational method used is also reflected in spatial distributions of charge and spin densities associated with the kinks. Along with the effects of electron-lattice interactions, in this paper we also study how the surrounding polar solvent influences the resulting structures, a much less explored topic. The solvent environments are ubiquitous in many applications involving fundamental redox processes. While neutral oligomers are practically not affected by the surrounding polar medium, the charged kinks are found to strongly interact with the solvent. We have recently demonstrated \cite{MGShort_prb,MGLong_jcp} at the level of \textit{ab initio} computations  that solvation of excess charge carriers on even-$N$ oligomers results in the formation of polaronic states, and that solvation and lattice distortions can act synergistically to increase the degree of charge localization. As the kink structure is present already on neutral odd-$N$ oligomers, one cannot characterize it as resulting from self-localization of an excess electron. We however observe that the spatial extent of the charged kink lattice distortion pattern gets shorter upon solvation, and in this sense the synergetic effect also takes place. This qualitative observation is independent of the computational method used, as is the fact that the solvent environment affects the spatial distribution of charge densities much more significantly than that of spin densities. This may be an indication of a different character of many-electron system responses to charge and spin perturbations.

\section{Computational methods and data processing}\label{systems}

All \textit{ab initio} computations in this study were performed using the
GAUSSIAN 03 suite of programs \cite{g03}. We have employed purely Hartree-Fock (HF) and DFT (PBE) calculations along
with the well-known \textit{hybrid}-DFT functionals BHandHLYP and B3LYP in order to compare
results of the different levels of theory. We remind the reader that the amounts of HF exchange included in these hybrid methods
are $50\%$ for BHandHLYP and $20\%$ for B3LYP. A rich 6-311++G($d,p$) all electron basis set has been employed
throughout. It should be noted that wherever possible, the results of our ``in vacuum" calculations have been
verified to compare well against previously published \textit{ab initio} data on even-$N$ \cite{yang_kertesz_2,schaefer_1} and odd-$N$  \cite{williams_1,schaefer_2} neutral and charged $\form$ systems both in terms of energetics and optimized
bond lengths. In order to investigate the effects
of solvation in a polar medium, GAUSSIAN 03 \cite{g03} offers its implementation
of the polarizable continuum model (PCM) described in original publications \cite{tomasi_1,
tomasi_2,tomasi_3,barone_1}. For all ``in solvent" PCM calculations, water has been chosen as a polar solvent
with its default parameters in GAUSSIAN 03.

In working with long oligomers we met with certain computational method-dependent limitations on achieving  satisfactory levels of convergence and spin contamination. In order to produce reliable results, therefore, we had to resort to using both restricted (RO, for HF and BHandHLYP) and unrestricted (U, for B3LYP and PBE) calculations \cite{Szabo}. The longest odd-$N$ oligomers studied have been $N=41$ (and some $N=61$) for HF, $N=61$ for BHandHLYP, $N=101$ for B3LYP, and $N=141$ for PBE methods. To our knowledge, these oligomer lengths are appreciably longer than used previously in comparable calculations.

The physical quantities of interest to us in this paper are the equilibrium (relaxed) geometries of relative carbon atom positions, atomic charge densities (to be denoted $\qn$) and atomic spin densities ($\rn$). We refer the reader to our previous publication \cite{MGLong_jcp} for an extensive discussion of some aspects of a retrieval of charge densities. The charge densities $\qn$ displayed below in Fig.~\ref{fig:bla_chrg} are \textit{excess} charges understood as the \textit{difference} of the raw charges in charged- and neutral-kink configurations. The raw atomic charges, in turn, are derived from L\"{o}wdin population analysis \cite{g03,leach_1}. We note that, for excess charge, the difference between L\"{o}wdin and Mulliken \cite{g03,leach_1} populations is relatively small \cite{MGLong_jcp}. For the atomic spin densities $\rn$, shown in Figs.~\ref{fig:SD_NEU} and \ref{fig:BLA_SD_MI}, we use the only available Mulliken population data.

A very convenient and well-known \cite{YuLubook,Barford_book,yang_kertesz_1,yang_kertesz_2} way to characterize the geometry of dimerized polymeric structures is via bond-length alternation (BLA) patterns
\begin{equation}\label{defd0}
\dn = (-1)^n \, \left(l_n - l_{n-1} \right),
\end{equation}
where $l_n$ is the length of the $n$th \textit{carbon-carbon} bond defined,  e.~g., as on the right of the $n$th carbon atom (we do not discuss the end bonds to hydrogens). For representation in this paper, we chose the index of the central carbon atom of the oligomer as $n=0$. In the infinite dimerized neutral structure, the dimerization pattern (\ref{defd0}) would be uniform, that is, independent of the spatial bond position $n$, we denote the \textit{magnitude} of this equilibrium ground-state pattern as $\dno$. This is not to be confused with a specific site value of $\delta_{n=0}$, the latter is not being explicitly used anywhere in the paper. The double degeneracy of the ground state of the infinite polymer is in that the uniform pattern (\ref{defd0}) can assume either value of $+\dno$ or $-\dno$ \cite{YuLubook,Barford_book}. For our purposes, we retrieve the magnitude $\dno$ from the middle segment of long even-$N$ oligomers featuring a well-established spatially-independent behavior.

\begin{figure}
\centering
\includegraphics[scale=0.65]{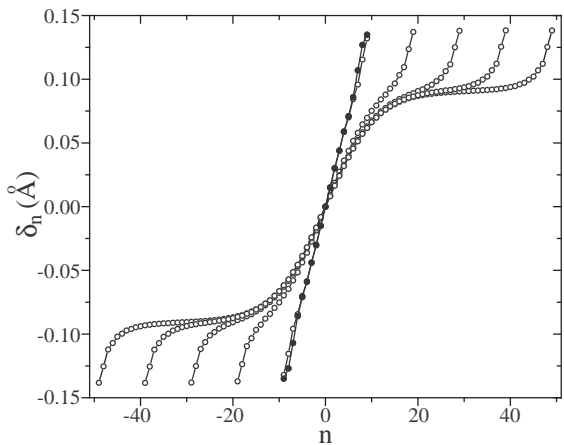}
\caption{BLA patterns for a series of increasing chain length neutral odd-$N$ oligomers (from $N=21$ to $N=101$)
computed using the B3LYP method, empty data points. Filled data points show results for the $\ctwentyone$ oligomer derived in Ref.~\cite{schaefer_2}.}
\label{fig:B3LYP_BLA}
\end{figure}
\begin{figure}
\centering
\includegraphics[scale=0.41]{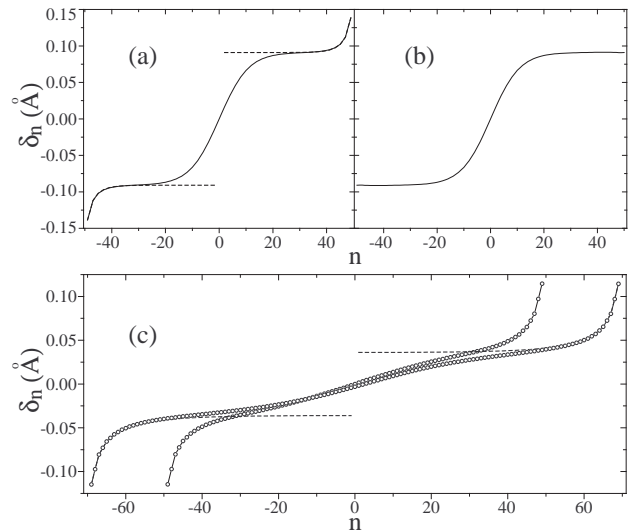}
\caption{Illustrating end effects in BLA patterns and their elimination: Panel (a) compares the neutral kink structure on $\chundredone$ (solid line) with the BLA patterns of ``halves'' of the neutral $\chundred$ oligomer (dashed), both obtained with the B3LYP method. After the subtraction of the common end BLA structures in this case, the kink BLA pattern of panel (b) results. Panel (c) compares relaxed BLA patterns on neutral $\chundredone$ and $\chundredfourtyone$ (data points) as well as ``halves'' of BLA patterns of the ground state of $\chundredfourty$ (dashed lines), all obtained with the PBE method.}
\label{fig:BL-BLA}
\end{figure}

As an example, Fig.~\ref{fig:B3LYP_BLA} displays the relaxed non-uniform BLA patterns for a series of neutral oligomers $\form$ ($N=21$, 41, 61, 81, and 101), all obtained within B3LYP computations. Along with our results also shown and in very good agreement is the data from Ref.~\cite{schaefer_2} on $\ctwentyone$ oligomer (the longest they studied was with $N=23$). The figure makes it transparent that the kink structure continues to evolve in longer oligomers well beyond $N=21$, and the latter case cannot serve even approximately to estimate the inherent extent of the kink. While illustrating the importance of a sufficient length of the oligomer, the figure also clearly shows that the end effects on the BLA pattern are quite substantial and persistent. The ``shape'' of the end-specific BLA structures stabilizes and becomes easily discernible in longer oligomers, it, of course, coincides with the one observed in BLA patterns of the ground state of even-$N$ oligomers. This fact allows to ``eliminate'' the end effects by comparison and an appropriate ``subtraction'' procedure, as illustrated in Fig.~\ref{fig:BL-BLA}(a) and (b). Such a procedure of removing the end effects from the finite oligomer data has been used to produced the kink BLA patterns shown in Fig.~\ref{fig:BLA_fit} and other figures in this paper.

\section{Neutral kinks}\label{sec:neutral_vac}

To quantify the spatial extent of kink-solitons displayed in in Fig.~\ref{fig:BLA_fit}, one can conveniently use the analytical shape of
\begin{equation}\label{sol}
\dn = \dno\, \tanh\frac{n}{l}
\end{equation}
well-known \cite{YuLubook,Barford_book} from the continuum electron-phonon models of CPs. Since the ground-state BLA magnitude $\dno$ has already been determined from the computation of long even-$N$ oligomers, the soliton half-width $l$ is the only parameter to be found from the fitting to the actual data. As $n$ in Eq.~(\ref{sol}) is the dimensionless index, parameter $l$ measures the spatial extent in units of the corresponding carbon-carbon lattice constants. The resulting fits for each of the computational methods used are shown in Fig.~\ref{fig:BLA_fit} and appear quite representative. The numerical values of the optimal parameters are collected in Table \ref{tbl1}. The table also
\begin{table}
\caption{\label{tbl1}Numerical values obtained for the dimerization magnitude $\dno$, neutral kink half-width $l$, and HOMO-LUMO gap $E_{g}$ with different methods from the longest oligomers studied.}
\begin{ruledtabular}
\begin{tabular}{l c c c}
method & $\delta_{0}$ (\AA) & $l$ & $E_{g}$~(eV)  \\
\hline
ROHF & 0.183 & 1.5 & 8.59\\
ROBHandHLYP & 0.135 & 3.5 & 4.18\\
UB3LYP & 0.091 & 10.8 & 1.77\\
UPBE & 0.036 & $\approx$ 29 & 0.41\\
\end{tabular}
\end{ruledtabular}
\end{table}
features the magnitude $E_g$ of the gap between the highest occupied (HOMO) and lowest unoccupied (LUMO) molecular orbitals of the neutral even-$N$ oligomers. The values presented in the table have been derived from the data on the longest oligomers studied: $l$ for the kink structures in Fig.~\ref{fig:BLA_fit}, $\dno$ and $E_g$ for $N=100$ oligomers in the case of HF, HandHLYP and B3LYP methods, and for $N=140$ in the case of PBE.

We should note here that a careful examination of the fitting to the PBE-derived kink BLA pattern on $\chundredfourtyone$ suggests that even on such a long oligomer, there is still some ``pressure'' exerted on the kink by the oligomer ends, and the derived value of $l \approx 29$ may be a bit underestimated. At the same time, there are clear indications that a fully-relaxed kink structure with a self-consistent spatial size can be found formed within a purely DFT-type computational framework. We consider this a useful observation bearing on the inherent qualitative capabilities present in such frameworks, even if the numerical magnitudes may be off the mark. In the other extreme of the HF computations, the resulting kink width (in agreement with earlier calculations \cite{williams_1}) is so small that the continuum-like analytical form (\ref{sol}) also has its limitations in representing some of the data points.

\begin{figure}
\centering
\includegraphics[scale=0.65]{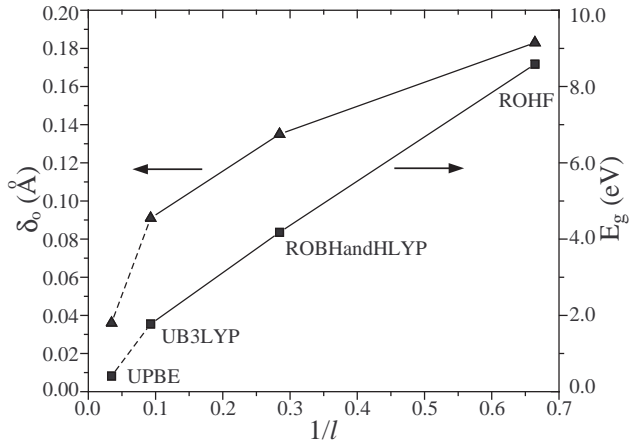}
\caption{The dimerization magnitude $\dno$ and HOMO-LUMO gap $E_{g}$ as functions of the inverse neutral kink half-width $1/l$ obtained with different methods, see Table \ref{tbl1}. Lines connecting data points are used here only to guide the eye.}
\label{fig:Eg_l}
\end{figure}
The overall picture of the relationships between results obtained with different methods is displayed in Fig.~\ref{fig:Eg_l} in the form of $\dno$ and $E_g$ as functions of the inverse kink half-width $1/l$. This view is motivated by the continuum electron-phonon models of CPs (\cite{YuLubook,Barford_book} and, particularly for polyynes, \cite{rice_2,rice_3}), where the kink energy gap parameter
\begin{equation}\label{CP1}
\Dx = \Do\,\tanh\frac{x}{\kso}
\end{equation}
as a function of the coordinate $x$ along the (infinite) polymer depends on the equilibrium magnitude $\Do$ and correlation length $\kso$ that are related as
\begin{equation}\label{CP2}
\kso = v_F/\Do
\end{equation}
($v_F$ being the Fermi velocity). More generally, relationships like Eqs.~(\ref{CP1}) and (\ref{CP2}) are also found for kink-solitons of nonlinear field theories, such as $\phi^4$  \cite{raja82}. The HOMO-LUMO gap $E_g$ in such models is just $2\Do$, and our dimensionless half-width $l \propto \kso$, hence the idea to see if the relationship
\begin{equation}\label{rel1}
E_g \propto 1/l
\end{equation}
might work for our data. In purely electron-phonon models (no electron-electron interactions), however, the magnitude of the gap parameter and the dimerization amplitude are just proportional to each other: $\Do \propto \dno$, and one therefore could as well be verifying the relationship
\begin{equation}\label{rel2}
\dno \propto 1/l.
\end{equation}

Both the dimerization amplitude $\dno$ and the kink spatial extent $l$ reflect the strength of the effective electron-phonon interaction: the stronger the interaction, the larger $\dno$ and the smaller $l$ are expected to be. The data presented in Table \ref{tbl1} and in Fig.~\ref{fig:Eg_l} clearly and consistently demonstrate that the effective electron-phonon interaction as revealed in results of \textit{ab initio} computations becomes stronger with the increased amount of HF exchange present in the energy functionals of the corresponding frameworks. Of course, the HOMO-LUMO gap $E_g$ also shows a steep growth with the amount of HF exchange. One should however recall that electron-electron interactions affect $E_g$ not only  ``indirectly'' via the effective magnitude of the electron-phonon interaction (and resulting dimerization $\dno$) but also ``directly''(as is easy to verify by keeping $\dno$ artificially fixed at non-optimal values). It is in this connection that it becomes interesting to compare our calculated data with both relationships (\ref{rel1}) and (\ref{rel2}), that would be equally valid in non-interacting electron models.

Already a cursory look at Fig.~\ref{fig:Eg_l} is sufficient to see that deviations from simple proportionality (\ref{rel2}) are quite substantial. On the other hand, the data is relatively much closer to relationship (\ref{rel1}), although by no means exactly following it. We cannot exclude that this interesting observation bears on the possibility of (rough) mapping between results of different models. One interpretation of the observation is that it is indeed the ``full'' energy gap parameter (\ref{CP1}) (rather than just the lattice displacements) that is a (more) adequate order parameter to describe spatially inhomogeneous structures within an effective nonlinear field model.

\begin{figure}
\centering
\includegraphics[scale=0.65]{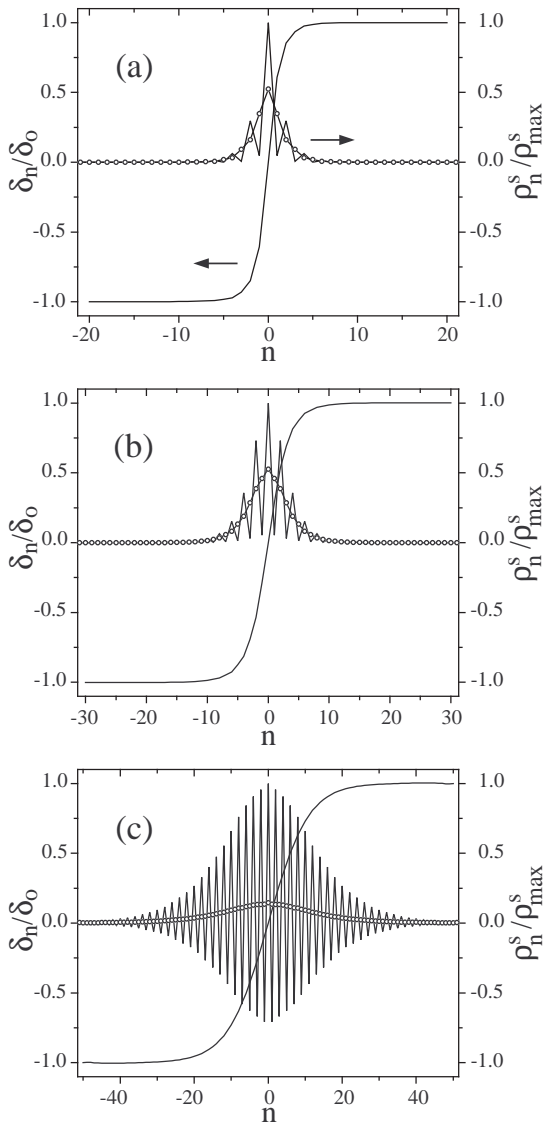}
\caption{The spatial distribution of BLA patterns and the spin density in neutral triplet kinks as derived with different methods: (a) ROHF, (b) ROHandHLYP, (c) UB3LYP. The results are shown in forms scaled to unit maximum values. The scaled raw (oscillating) atomic spin densities are displayed as simple solid lines, their averaged (as described in the text) behavior is indicated with data symbols.}
\label{fig:SD_NEU}
\end{figure}

The polyynic structures, in comparison with polyacetylene, possess an extra degeneracy of $\pi$-electron single-electron levels, resulting from two independent orientations of wave functions in the plane perpendicular to the polymer axis. The midgap electronic level formed in the presence of the kink-shaped BLA pattern has therefore an internal degeneracy of four (including spin) and is occupied by two electrons when the kink is uncharged \cite{rice_2,rice_3}. The neutral kink therefore can be either in the singlet or triplet states. In the non-interacting electron models, these states are degenerate. It is known though from the previous studies \cite{williams_1,schaefer_2} that electron interactions prefer the triplet state. In full agreement, all our simulations have also resulted in the triplet, spin $S=1$, ground state of neutral kinks. Only these triplet states are being explicitly presented in this paper. In the context of our discussion, this comes in ``handy'' as one can compare the spatial structures of BLA patterns with non-vanishing spin density distributions.

To facilitate this comparison, in Fig.~\ref{fig:SD_NEU} we plot BLA patterns and spin densities (retrieved from the Mulliken population analysis) in the scaled form emphasizing their spatial structures. (From now on, comparison results would be shown as derived with three methods excluding PBE.) The raw atomic densities exhibit characteristic oscillations from carbon to the next carbon. To better see the spatial extent of the spin density, Fig.~\ref{fig:SD_NEU} also displays the averaged behavior of scaled atomic densities, where a simple-minded nearest-neighbor averaging of site-$n$ specific quantities $r_n$ is done as $\langle r_n \rangle = 0.5 (r_n + 0.5 (r_{n-1}+r_{n+1}))$. (We use the same type of averaging for data in Figs.~\ref{fig:bla_chrg} and \ref{fig:BLA_SD_MI}.) It is well known \cite{leach_1} that calculations of atomic charges from \textit{ab initio} results depend on the basis set transformations and sometimes lead to artifacts and spurious results. While we cannot exclude that some artifacts may be reflected in the raw atomic-centric data, we expect the averaged behavior to be much less susceptible to such dependencies (compare to our discussion \cite{MGLong_jcp} of atomic- and cell-centric charge densities). With the averaged data, it is especially transparent from Fig.~\ref{fig:SD_NEU} that for all computational methods used, there is a very good correspondence of the spatial extent of the kink BLA structure and the region of the spin localization. Needless to emphasize that, of course, each of the methods yields a very different spatial size of this region. In terms of the magnitude of oscillations of atomic densities, the panel (c) of course stands out, this may be related to the fact that the data for that panel was generated with an unrestricted computational method, while panels (a) and (b) have been produced with restricted computations.

As was expected from our earlier analysis \cite{MGShort_prb,MGLong_jcp} of neutral even-$N$ oligomers, we have found practically no effect produced by solvation on the properties of neutral kinks at all levels of computation. That is why the data shown for neutral kinks has been restricted to results ``in vacuum'' only. The situation changes substantially for charged kinks.

\section{Charged kinks and the effects of solvation}\label{sec:charged_solvent}

\begin{figure*}
\centering
\includegraphics[scale=0.8]{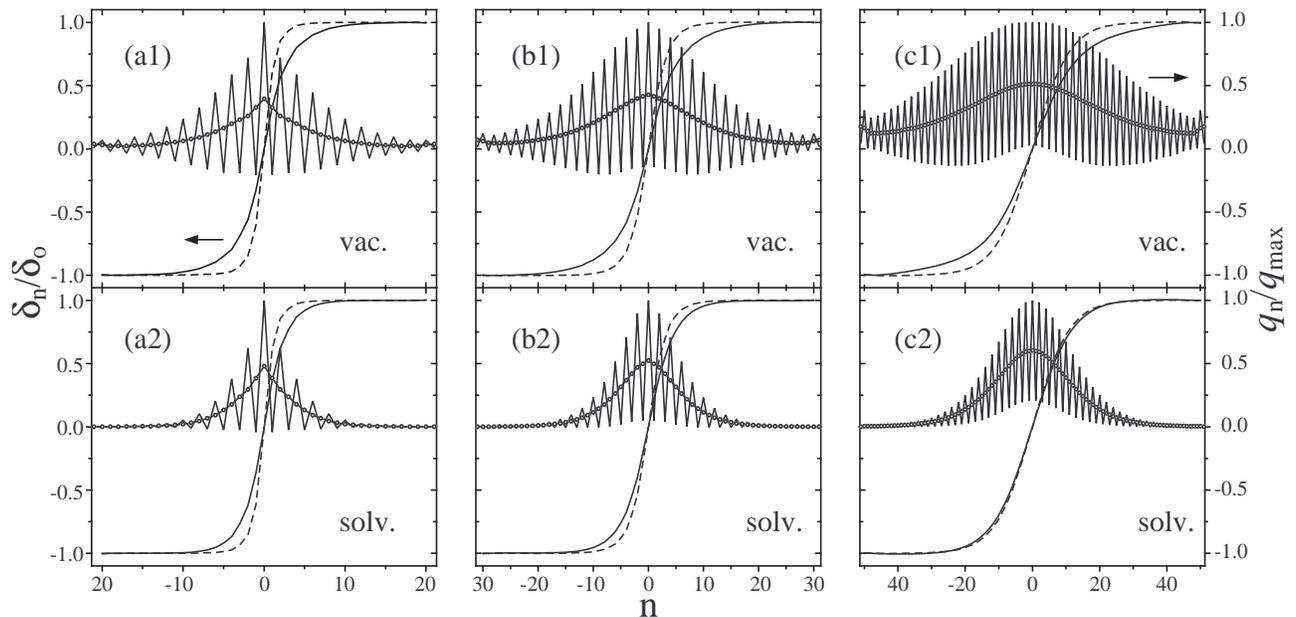}
\caption{BLA patterns and excess charge distributions for singly-charged kinks as obtained with different methods:  ROHF in columns (a), ROHandHLYP in (b), and UB3LYP in (c). The first row displays results for chains in vacuum and the second for chains in the solvent environment. To compare, BLA structures for neutral kinks are also depicted with dotted lines. The raw atomic charge densities scaled to the unit maximum values are shown by simple solid lines, their nearest-neighbor-averaged (as discussed in Sec.~\ref{sec:neutral_vac}) behavior is by lines with data symbols.}
\label{fig:bla_chrg}
\end{figure*}
\begin{figure*}
\centering
\includegraphics[scale=0.8]{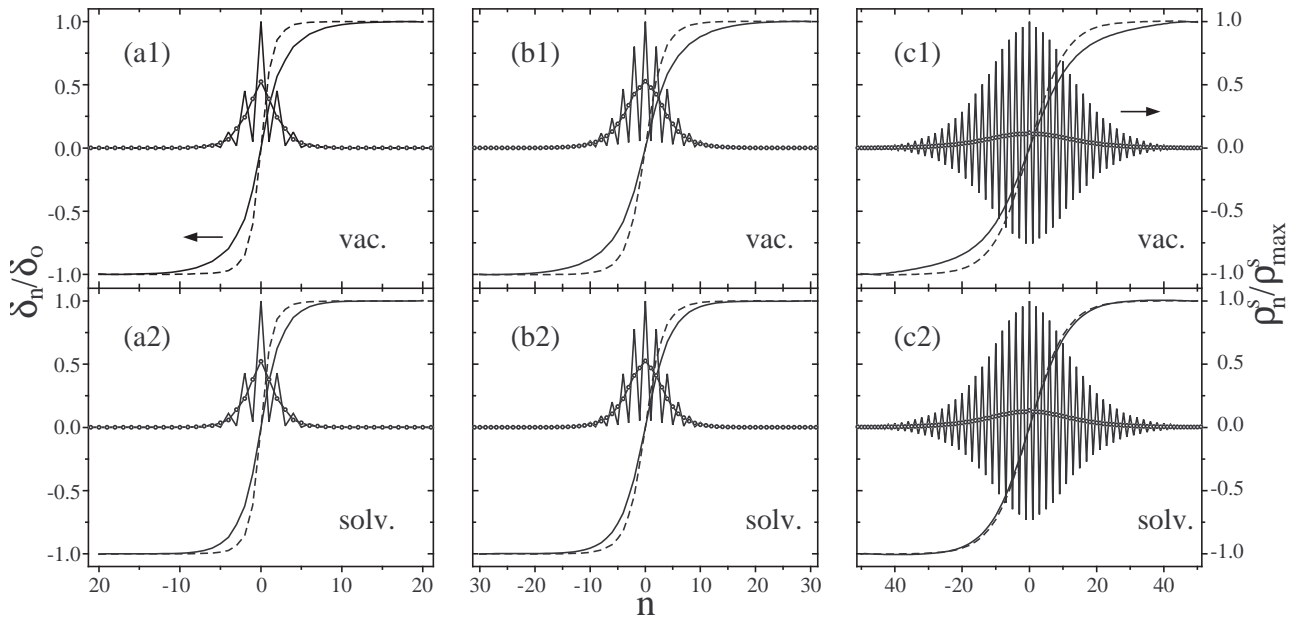}
\caption{As in Fig.~\ref{fig:bla_chrg} but for spin (instead of charge) density distributions.}
\label{fig:BLA_SD_MI}
\end{figure*}

In this Section we discuss characteristics of singly-charged kinks resulting upon the addition of an extra electron to long odd-$N$ polyynic oligomers. (These systems and nearly charge-conjugation symmetric and similar results are obtained upon the addition of an extra hole \cite{MGLong_jcp}.) Our discussion will revolve around the data shown in Figs.~\ref{fig:bla_chrg} and \ref{fig:BLA_SD_MI} displaying resulting equilibrium distributions of BLA patterns, excess charge and spin densities. The overall charge of the system is of course just the electronic charge and the spin $S=1/2$. While different computational methods yield numerically different magnitudes, common qualitative features are clearly discernible in the figures.

All results indicate localized characteristics around the center of the kink in the middle of oligomers. This should not be construed as self-localization of an excess electron since midgap levels are formed already within the neutral kink structure and can be readily occupied by that electron.  As perceived via BLA patterns, charged kinks evidently become wider in their spatial extent than neutral kinks, which would not be happening in non-interacting electron models \cite{rice_2,rice_3}.  This appears natural to rationalize as the effect of Coulomb forces trying to spread out the excess charge, as if there was an effective width-dependent capacitance associated with the kink. This view is further supported by the observation of an appreciable effect of the screening solvent environment on the spatial extent of the kink. As is clearly seen, solvation decreases the extent of charged kinks in all computations and practically restores the neutral kink BLA pattern in the case of B3LYP (columns (c)).

We did not attempt to analyze the length of charged kinks by fitting to the analytical dependence (\ref{sol}) as even longer oligomers would need to be studied to avoid the influence of end effects. In addition to the visual perception of those in the figures, the data in Table \ref{tbl2} illustrates oligomer-length-dependent progression of the electron affinities $\EA$ understood as the difference of the total (relaxed) system energies of neutral and singly-charged kinks:
$$
\EA=\Etot^{0}-\Etot^{-}.
$$
\begin{table}
\caption{\label{tbl2}Electron affinities in eV calculated with various methods in vacuum and in the solvent for singly-charged oligomers of different lengths.}
\begin{ruledtabular}
\begin{tabular}{l c c c c c c}
method &\multicolumn{2}{c}{$\cfourtyone$}&\multicolumn{2}{c}{$\csixtyone$}
&\multicolumn{2}{c}{$\chundredone$}\\
& vac& solv&vac&solv&vac&solv\\
\hline
 & 2.39 & & 2.43 & &  & \\
\raisebox{1.5ex}[0cm][0cm]{ROHF} & & 3.66 & & 3.66 & &  \\
\hline
 & 3.69& & 3.78 & &  & \\
\raisebox{1.5ex}[0cm][0cm]{ROBHandHLYP} & & 4.68 & & 4.69 & & \\
\hline
 & 3.88& & 4.09 & & 4.23 & \\
\raisebox{1.5ex}[0cm][0cm]{UB3LYP} & & 4.68 & & 4.72 & & 4.73 \\
\end{tabular}
\end{ruledtabular}
\end{table}
One can see that this quantity in the vacuum data has not converged yet to self-consistent (length-independent) values. In contrast, the convergence has been (nearly) achieved for the ``in solvent'' data. That corresponds very well with the picture of a great effect of the solvent on the spread of the excess charge density in Fig.~\ref{fig:bla_chrg} which is obviously much farther away from oligomer ends in the solvated case (second row). Comparing $\EA$s in vacuum and in solvent, the table also shows the energetic significance of solvation: from more than 1.2 eV in HF computations to 0.5 eV in B3LYP. The method dependence here is consistent with the fact that the spatial extent of the excess charge is shortest in HF, hence the electrostatic Coulomb effects are strongest.

A remarkable and consistent difference is transparent in the behavior of charge and spin densities in Figs.~\ref{fig:bla_chrg} and \ref{fig:BLA_SD_MI} for all computational methods. The spatial spread of the charge densities in vacuum is much more significant than of the spin densities. The former strongly respond to the solvent environment, while there is very little effect on the latter. Even in the solvent the charge and spin densities appear perceptibly distinct. This difference is at variance with what one would expect in a simple picture of a single localized electron, whose probability density would establish both charge and spin densities. Evidently, it is the response of the many-electron system that should account for the observed difference.

Given the fact that the solvent so strongly affects the excess charge distribution, one could relate the effect to long-range electrostatic Coulomb forces that polarize the many-electron system of the polyyne and cause remote variations of the charge density but not the spin density. This however does not appear as a complete explanation as the difference is observed for the averaged densities as well (``true'' excess charge). Apparently, there may be inherent distinctions in system responses here to charge and spin perturbations, perhaps related to the nature of the order parameter. While, of course, not directly applicable to our system, one cannot help speculative wondering if there could be some relationship to the phenomenon of spin-charge separation well-known in 1D electronic liquids \cite{GV2005}.

At the same time, we recognize that some computational artifacts may have affected the observed picture. We note that the overall spatial behavior of spin densities on charged kinks is similar to observed on neutral kinks, Fig.~\ref{fig:SD_NEU}, and one is reminded that it is only B3LYP computations (column (c) in Figs.~\ref{fig:bla_chrg} and \ref{fig:BLA_SD_MI}) that have been performed in the unrestricted mode. This may have contributed to larger degrees of spatial variations of the atomic spin densities observed in that case. It is not clear if more significant spin responses would have resulted from unrestricted calculations with HF and HandHLYP functionals.

\section{Discussion}

Structural simplicity of polyynic carbon chains $\form$ makes them a good prototypical example of 1D semiconductors and a convenient ``playground'' for using various theoretical methods. In this paper we extended the previous work on these systems to systematically study characteristics of kink-solitons that are naturally formed in bond-length alternation patterns of odd-$N$ oligomers. We attempted a quantitative comparison of results that are obtained at different well-known levels of \textit{ab initio} computations: from pure DFT (PBE) via hybrid B3LYP and BHandHLYP to pure Hartree-Fock (HF). These methods have been listed here in the order of the increased amount of HF exchange in their functionals (0, 20, 50, and 100\%, respectively). The earlier findings of the increased magnitude of the uniform lattice dimerization pattern ($\dno$) and the HOMO-LUMO gap ($E_g$) with the increased amount of HF have now been complemented by explicit observations of the decreased spatial extent ($l$) of the kinks. This is consistent with  stronger effective electron-phonon couplings ``induced'' by HF exchange. By studying long oligomers, we tried to minimize the influence of the end effects and to establish self-consistent sizes $l$ of neutral kinks as determined by inherent system interactions. While numerical values of resulting parameters vary substantially from method to method, the data suggests that the relationship $E_g \propto 1/l$ could serve as a rough first approximation for mapping results between different methods. This may be indicative of the energy gap parameter (including effects from both electron-phonon and electron-electron interactions) being an adequate order parameter for building a unified effective nonlinear field model.

The more important issue, however, is if any (single) of these methods is adequate to describe real systems, which cannot be resolved without verification against experiments. Based on comparisons with structural and optical experimental data, it was suggested previously \cite{yang_kertesz_1} that BHandHLYP may be the ``right'' method to compute the equilibrium lattice configurations of uniform systems but B3LYP should be used for the electronic structure.  The latter conclusion was motivated by BHandHLYP-derived HOMO-LUMO gaps $E_g$ being larger than experimental  ``optical gaps'' and would be unfortunate in the sense of the internal consistency of a single theoretical method. In view of the strong excitonic effects in 1D systems \cite{StandardModel,haugbook}, however, the situation may need to be reassessed to include those effects in calculations of the optical spectra. On the other hand, the problem of small DFT-derived energy gaps $E_g$ has been successfully remedied \cite{StandardModel} by calculating corrections to the quasi-particle energies within the framework of the GW-approximation \cite{Martin,StandardModel,GV2005} (excitons are further calculated via the Bethe-Salpeter equation \cite{StandardModel}). Even from a purely theoretical viewpoint, it would be quite interesting to compare results derivable for polyynes with the help of GW to the ones obtained with less computationally intensive hybrid-DFT methods used in this paper. The picture emerged in our discussion of various computational outputs (Fig.~\ref{fig:Eg_l}) strongly suggests, however, that they are intimately related, and just taking into account GW-corrections for $E_g$ may turn out not to be sufficient, it is the optimal lattice structure of the ground state itself that could undergo changes if treated on the same footing. Where in Fig.~\ref{fig:Eg_l} would then the corresponding data points fall appears a challenging question to answer.

Consistently with our previous studies \cite{MGShort_prb,MGLong_jcp} of polaronic states and qualitatively method-independent, we have found a substantial localizing effect that the surrounding solvent produces on the excess charge densities associated with charged kinks. As a result of solvation, the equilibrium spatial extent of charged kinks also becomes shorter; one may say that the lattice deformation and polarization of the solvent act synergistically to increase the degree of charge localization. We expect the drag from the solvent to significantly reduce the mobility of charged kinks as it does for solvation-induced polarons \cite{CBDrag,GU_lowfreq}. Quite different responses in charge and spin densities found in all computations clearly indicate that electron-electron interactions are strongly involved in accommodation of an excess charge carrier by the kink-soliton structure.

\bibliography{abinitio_ref}
\end{document}